\documentclass[12pt,preprint]{aastex}

\slugcomment{To be published in the Astrophysical Journal volume 622 on 1 April 2005}

\begin{document}
\title{GAMMA-RAY SPECTRAL STATE TRANSITIONS OF GRO~J$\bf1719-24$}
\author{J. C. Ling }
\affil{Jet Propulsion Laboratory 169-327, California Institute of Technology}
\affil{4800 Oak Grove Drive, Pasadena, CA 91109}
\email{james.c.ling@jpl.nasa.gov}
\and
\author{Wm. A. Wheaton }
\affil{Infrared Processing and Analysis Center, California Institute 
of Technology}
\affil{100-22, Pasadena, CA 91125}
\email{waw@ipac.caltech.edu }

\begin{abstract}

We report results of an in-depth study of the long-term soft $\gamma$-ray (30 keV -- 1.7 MeV) flux and spectral variability of the transient source GRO~J$1719-24$ that was first discovered by BATSE and SIGMA in the fall of 1993. Our results were obtained from the JPL BATSE-EBOP database covering a 1000-day period between 13 January 1993 and 10 October 1995. During this period, the source underwent a major outburst in the fall of 1993 when the 35-100 keV flux rose from a quiescent state of less than 16 mCrab before 17 September 1993 to a level of 1.5 Crab on 3 October. The source remained in this high-intensity state over the next $\sim$70 days during which the $35-100$ keV flux decreased monotonically by $\sim$33$\%$ to $\sim$1 Crab level on 12 December, then decreased sharply to the pre-transition quiescent level of $\sim$44 mCrab on 21 December where it remained until 5 September 1994. During a 400-day period between 5 September 1994 and 10 October 1995, the source again underwent a series of five transitions when the 35-100 keV flux increased to low-intensity levels of $\sim$200-400 mCrab, a factor of $4-7$ times lower than that observed in 1993. The low and high-intensity states were characterized by two different spectral shapes. The low-state spectra were described by a power law with spectral index of $\sim$2. The high-state spectra on the other hand have two components: a thermal Comptonized shape below $\sim$200 keV with electron temperature $kT_e$ of $\sim$37 keV and optical depth $\tau$ $\sim$2.8, and a soft power-law tail with photon index of $\sim$3.4 above 200 keV that extends to $\sim$500 keV. The softer high-intensity spectrum and the harder low-intensity spectrum intercept at $\sim$400 keV. The non-thermal power-law gamma-ray component in both the high and low-intensity spectra suggests that the persistent non-thermal emission source is coupled to the hot and variable thermal emission source in the system. Furthermore, the correlation of the spectral characteristics with the high and low-intensity state resembles that seen in two other gamma-ray emitting black-hole candidates GRO~J0422+32 and Cygnus X-1, suggesting that perhaps similar system configurations and processes are occurring in these systems. Possible scenarios for interpreting these behaviors are discussed.
  
\end{abstract}

\keywords{gamma-rays observations---Black Holes ---individual (GRO~J$1719-24$)}

\section{INTRODUCTION}

The transient gamma-ray source GRO~J$1719-24$ (Nova Oph 1993) was first discovered by BATSE on the Compton Gamma-Ray Observatory (CGRO; Harmon et al. 1993a) and SIGMA on Granat (Ballet et al. 1993) on September 25, 1993 [TJD 9255 where TJD (Truncated Julian Day)= JD (Julian Day) - 2,440,000.5] when the hard x-ray ($40-150$ keV) flux emerged suddenly from the quiescent state of 12 mCrab 2$\sigma$ upper limit measured by SIGMA (Revnivtsev et al. 1998), integrated over the period from 28 August to 19 September 1993 (TJD $9227-9249$),  to 17 mCrab on $25-26$ September 1993 (TJD $9255-9256$), and then to a peak level of 1.4 Crab measured by BATSE (Harmon et al. 1993b) five days later on September 30 (TJD 9260). BATSE (Harmon et al. 1993b) then showed that the hard x-ray flux decreased gradually by $\sim$0.3$\%$ $\pm$ 0.05$\%$ per day for the next two and half months to the level $\sim$1.1 Crab on 9 December (TJD 9330) before dropping off sharply in the next six days to below the BATSE 1-day sensitivity level at around December 16 (TJD 9337). The source stayed at this quiescent state until the fall of 1994 when SIGMA (Churazov et al. 1994; Revnivtsev et al. 1998) detected the hard x-ray flux again at $\sim$115 mCrab, $\sim$10$\%$ of the peak flux level observed in the fall of 1993. Strong X-ray emission was also detected by the Mir/Kvant experiment (Borozdin, Alexandrovich, $\&$ Sunyaev 1995) on 1995 February 16 (TJD 9764) and by BATSE (Hjellming et al. 1996) in five separate episodes between 1 September 1994 (TJD 9596) and 6 September 1995 (TJD 9995) when the hard x-ray flux reached a level of $\sim200-300$ mCrab (refer to Section 3 of this paper), a factor of $3-5$ lower that observed in 1993. 

The optical counterpart of GRO~J$1719-24$ was identified by Della Valle, Mirabel $\&$ Rodriguez (1994) and Masetti et al. (1996) to be a low-mass binary (V2293 Oph) with a periodicity of 14.7 hours. According to the optical superhump modulation analysis of Masetti et al. (1996), the system consists of a compact object of  $\>$4.9 solar mass, and a companion main sequence star of K spectral type (or later) of $\sim$1.6 solar mass. The distance was estimated to be $\sim2-2.8$ kpc. When the source was in the quiescent state prior to the transition, no optical counterpart was observed at a level fainter than $V\sim21$. During the 1993 outburst, the visual magnitude increased by $\sim$4.4 to $V=16.65$. A strong QPO peak in the 20-100 keV band was observed by BATSE (van der Hooft et al. 1996) during the $\sim$80-day high-intensity state in 1993 with centroid frequency varied from $\sim$0.04 at the onset of the outburst to $\sim$0.3 Hz at the end.  

The radio counterpart was observed by the Very Large Array (VLA) (Mirabel et al. 1993; Della Valle, Mirabel and Rodriguez 1994) during the 1993 outburst on 5 October at a position consistent with the optical counterpart. The radio compact source showed a relatively flat spectrum with flux densities of 4.6$\pm$0.4 and 4.9$\pm$0.2 mJy at $\lambda$20 cm and $\lambda$6 cm, respectively. Radio emission was also measured by the Molonglo Observatory Synthesis Telescope (MOST) at 843 MHz and by the VLA at frequencies of 1.49, 4.9, 8.4 and 14.9 GHz during the 1994-1995 (TJD 9745-TJD 9966) period when the source underwent a series of low-level outbursts described above (Hjellming et al. 1996). Specifically, the 4.9 Ghz measured by VLA between TJD 9749 and TJD 9800 showed a power-law decay of the lightcurve that matched closely to the decay of one of the events observed by BATSE. This was interpreted as the radio synchrotron emission associated with the ejection of relativistic electrons from an expanding spherical shell

Soft gamma-ray spectra of GRO~J$1719-24$ above 30 keV between January 1993 and September 1995 have been measured by both the SIGMA experiment (Revnivtsev et al. 1998) and by OSSE on CGRO (Grove et al. 1998). Revnivtsev et al. (1998) reported near contemporaneous observations with TTM onboard the MIR-KVANT and SIGMA in the composite 2-300 keV range made during seven separate periods: (a) two during the pre-outburst period, 17 February-9 April 1993 and 28 August-19 September 1993, (b) two during the rising phase ($25-28$ September 1993) and the plateau phase (29 September - 14 October 1993) of the 1993 outburst, and (c) three during the post outburst low-intensity periods, 24 February-25 March 1994, $1-30$ September 1994 and $9-21$ September 1995, respectively.  Grove et al. (1998) reported OSSE observations made in five separate periods, $25-27$ October 1993, $30-31$ October 1993, $9-15$ November 1994, 29 November- 7 December 1994 and $1-14$ February 1995. Both the SIGMA and OSSE results indicated possible variability between spectra obtained during the ``plateau" phase of the flare observed in the fall of 1993 and those obtained in the fall of 1994 when the flux was significantly lower. The 1993 high-intensity  spectra were generally better fitted with either the Comptonized model (Sunyaev $\&$ Titarchuk 1980), a thermal bremsstrahlung model or a exponentially truncated power law than a power law, while the 1994 low-intensity spectra were better fitted with either a power law or with a less well constrained thermal bremsstrahlung model. Grove et al. (1998) further suggested that of the seven galactic gamma-ray black-hole emitters observed by OSSE, the two spectral states observed in GRO~J$1719-24$ strongly resemble those seen in Cygnus X-1. 

The spectral variability shown between the high and low-intensity spectra observed by SIGMA and OSSE with limited pointed observations has important implications in our understanding of the system. It therefore needs to be confirmed and studied with all available data on the source. In this paper, we report results obtained from BATSE's near continuous daily monitoring of the $35-1700$ keV emission of the source covering a 1000-day period between TJD 9000 and 10000. These results, which were produced by the JPL EBOP (Enhanced BATSE Occultation package) analysis system (Ling et al. 1996, 2000), consist of lightcurves in six broad energy bands, $35-100$ keV, $100-200$ keV, $200-300$ keV, $300-400$ keV, $400-700$ keV and $700-1000$ keV, respectively, and 14-channel daily spectra as well as spectra integrated over days and weeks. In this paper, we will address the following questions: (1) What new information is revealed from the energy-dependent gamma-ray flux history and variability shown by this source during this period? (2) What are the characteristic of the high-intensity and the low-intensity soft $\gamma$-ray spectra in this energy range? (3) How did the source spectrum evolve between the low and high-intensity states? (4) How do the spectral characteristics of this source compare with other black-hole candidates such as GRO~J0422+32, Cygnus X-1, GRO~J$1655-40$ and GRS~1915+105 etc? and (5) What can we learn from the gamma-ray spectral characteristics and variability shown in these sources to better understand their system configurations and the physical processes? We address these issues in Section 4. In Section 2, we provide a brief description of the EBOP database and technique. Results produced by EBOP are presented in Section 3.

\section{EBOP Database and Technique} 

EBOP, the JPL Enhanced BATSE Occultation Package as described by Ling et al. (1996; 2000) and Ling $\&$ Wheaton (2003a,b) was used to produce the results reported in this paper. We have also completed recently a high level database with EBOP for studying the long-term flux and spectral characteristics and variations of 75 $\gamma$-ray sources included in the present EBOP catalog covering the full nine-year CGRO mission from 1991 to 2000. This high-level database includes daily count-rate and photon spectra (assuming a power-law fit, or fits with a Compton model whenever it is needed, to the count-rate spectrum using XSPEC [Arnaud 1996]), and spectra integrated over each of the ~315 nominal CGRO Viewing Periods (VP) of typically 6 to 14-day duration, for each of the eight BATSE Large Area Detectors (LADs) and for each of the 75 sources. Many terms in the background differ strongly for different LADs. The detector responses to any given source, which implicitly transform LAD count rates into photon fluxes, also differ. To reduce the possible effects of systematic errors in the results, EBOP includes a LAD flux consistency check. Days when the fluxes derived from the source-viewing LADs (typically two to four for each VP) are mutually inconsistent (at the 95\% level) are rejected from inclusion in the results. Because some terms
in the background (direct cosmic-ray effects, in particular) are correlated with the earth-occultation periodicity, this filter on the results is not truly airtight, though logically necessary and useful. Of the 780 daily spectra covering the period between TJD 9000 and 10000 included in this report, only two have failed the consistency test, using the criteria described by Ling et al. (1996, 2000). 

\section{Results}
\subsection{Flux Histories}

The flux histories in the six broad-energy bands, $35-100$ keV, $100-200$ keV, $200-300$ keV, $300-400$ keV, $400-700$ keV, and $700-1000$ keV, covering the period between 13 January 1993 (TJD 9000) 
and 10 October 1995 (TJD 10000) are shown in Figure 1. The resolutions are one day for the first two panels, two days for the 3rd panel, and five days for the last three panels. Between TJD 9264 and 9341, CGRO underwent three reboost operations on TJD 9264 and 9279, TJD 9310-9314, and TJD 9334-9341, respectively, during which no data on the source were available as reflected by the data gaps shown in these lightcurves.

Over the 1000-day period, the source underwent a series of gamma-ray flux transitions. 

1. The hard x-ray (35-100 keV) flux rose sharply from the quiescent state (Q1) of (2.19 $\pm$ 0.74) $\cdot 10^{-5}$ photons cm$^{-2}$-s$^{-1}$-keV$^{-1}$ ($\sim$10 mCrab) measured between TJD 9000 and TJD 9247 to 114 mCrab on TJD 9249 ($\sim$2$\sigma$), the first hint of the beginning of a transition, and then to $\sim$1.5 Crab ($\sim$34$\sigma$) 14 days later on TJD 9263. All errors associated with the measured fluxes reported in this paper are 1$\sigma$ significance.
 
2. The source stayed roughly in the high-intensity state for $\sim$76 days between TJD 9257 and 9333. During this period, the 35-100 keV flux decreased gradually from its peak value on TJD 9263 of 1.5 Crab to $\sim$1 Crab on TJD 9333 before the 3rd reboost. When we observed the source again at the other end of the reboost on TJD 9342, the source flux returned roughly to the pre-flare level (Q2). The averaged 35-100 keV flux over the next $\sim$260 days from TJD 9342-9600 was (9.28 $\pm$ 0.88) $\cdot 10^{-5}$ ($\sim$44 mCrab) photons cm$^{-2}$-s$^{-1}$-keV$^{-1}$.

3. During the high-intensity period between TJD 9262 and TJD 9333, BATSE observed high-energy gamma-ray fluxes up to $\sim$400-700 keV at the level of (4.12 $\pm$ 1.29) $\cdot 10^{-6}$ ($\sim$0.28 Crab). Table 1 lists the average fluxes measured in the six energy bins during the high-intensity period as well as the quiescent periods before and after the outburst. The average 35-100 keV flux measured during the high state was $\sim$1.2 Crab at the level of (2.57 $\pm$ 0.02) $\cdot 10^{-3}$ photons cm$^{-2}$-s$^{-1}$-keV$^{-1}$. 

4. During the 400-day period between TJD 9600 and 10000, the source flared again five different times (marked as ``14", ``15", ``16", ``17" and ``18" in the first panel of Figure 1) but to levels significantly lower than the high-state level. The average 35-100 keV flux measured in these five periods are (7.38 $\pm$ 0.27) $\cdot 10^{-4}$ ($\sim$354 mCrab), (7.40 $\pm$ 0.27) $\cdot 10^{-4}$ ($\sim$354 mCrab), (4.12 $\pm$ 0.26) $\cdot 10^{-4}$ ($\sim$197 mCrab), (3.95 $\pm$ 0.21) $\cdot 10^{-4}$ ($\sim$189 mCrab), and (6.05 $\pm$ 0.31) $\cdot 10^{-4}$ ($\sim$290 mCrab) photons cm$^{-2}$-s$^{-1}$-keV$^{-1}$, respectively. Gamma-ray fluxes for the six broad-band energy-bins for each of these five periods are summarized in Table 2.

5. During this entire 1000-day period, GRO~J$1719-24$ was also observed in six separate periods of various duration by SIGMA (Revnivtsev et al. 1998) and five times by OSSE (Grove et al. 1998). These are identified as ``a"$-$``f" and ``A"$-$``E", respectively, in panel 2 of Figure 1. We include these results for the purpose of comparing them with the BATSE results in the context of the source's long-term flux and spectral variability observed by BATSE. Specifically, the BATSE data as well as those obtained by SIGMA and OSSE are essentially the world's only available data on this source for shedding light on the high-energy processes as well as, possibly, its system configuration as revealed by the gamma-ray data. 

\subsection{Spectra}

The key questions we address in this section are: what are the basic spectral characteristics associated with the various $\gamma$-ray states, and how do they evolve as the source underwent transitions among these states? 

1. Figure 2a shows nine daily spectra measured between TJD 9249 and TJD 9260 during the rising phase of the 1993 source transition from the quiescent state (Q1) to the high state. In each panel, the solid line is the best-fit model to the spectral data measured by all the ``source-viewing" LADs listed in Table 3. Pertinent information of the model fit (e.g. best-fit parameters with 1$\sigma$ error) as well as reduced $\chi^2$ ($\chi^2$/$\nu$, where $\nu$ is the number of degrees of freedom) is also displayed in each panel and listed in Table 3. 

Due to limited statistics of the data, the first five spectra measured between TJD 9249 and TJD 9256, can be adequately fitted with either the power law or the thermal Comptonization model (Sunyaev \& Titarchuk 1980). However, on TJD 9257, the spectrum has a distinctly breaking ``thermal" shape, and the simple power law can no longer fit the data well. Since the integrated 35-500 keV flux changed by only $\sim$18$\%$ from $\sim$932 mCrab on TJD 9256 to $\sim$1.1 Crab on TJD 9257, a reasonable question may be raised, ``Was there a definitive spectral change from either a non-thermal power-law shape or a breaking thermal shape seen prior to TJD 9257 to the thermal Comptonization shape on TJD 9257 and beyond?" Such a change could reflect an intrinsic change of the physical conditions of the system, and needs to be understood. In addressing this question, we examine the hardness ratio (HR) of the spectra, where HR = $150-300$ keV flux/ $35-150$ keV flux (see Table 3). We then use HR to characterize the broad spectral shape of the low and high-intensity spectra measured during both the 1993 transition (Figure 2a spectra 1-9, and Figure 2b spectra 10-11) as well as those measured in 1994-1995 (Figure 2b spectra 12-18). For TJD 9255 and 9256, the HRs were measured to be (8.68 $\pm$ 2.16) $\cdot 10^{-2}$ and (7.21 $\pm$ 1.08) $\cdot 10^{-2}$, respectively. The weighted average of the two values is (7.58 $\pm$ 0.96) $\cdot 10^{-2}$. This is compared to (5.07 $\pm$ 0.86) $\cdot 10^{-2}$ for the TJD 9257 spectrum. The difference is $\sim$2$\sigma$ significant. Since the average HR of (7.58 $\pm$ 0.96) $\cdot 10^{-2}$ measured on TJD 9255 and 9256 is consistent with the average HR of the seven low-intensity power-law spectra measured in 1994-1995 (Figure 2b spectra 12-18) of (7.96 $\pm$ 0.39) $\cdot 10^{-2}$, these results suggest that the source spectrum changed from a power-law to a thermal Comptonization shape within one day from TJD 9256 to TJD 9257 during the rising phase of the 1993 transition. 

2. The source spectrum remained in this same Comptonized shape for the rest of the rising phase (Figure 2a panels 6-9) from TJD 9257 to TJD 9260 and when the source was in the high intensity state from TJD 9262 to TJD 9333 (Figure 2b panels 10-11). The weighted average of the hardness ratio of these six spectra is (5.61 $\pm$ 0.16) $\cdot 10^{-2}$.

3. The two averaged high-state spectra with improved statistics integrated over TJD $9262-9299$ and TJD $9301-9333$ (Figure 2b panels 10 and 11), respectively, and their combined spectrum shown in Figure 3 specifically, showed two components: a soft power-law tail of photon index of 3.37 above 200 keV, that extended to $\sim$500 keV, superposed on the low-energy Comptonized component below 200 keV. Because the declining phase from the high state to the quiescent state occurred during the period of the reboost of the CGRO spacecraft, no information on the temporal structure of the decline, the date and the flux threshold at which the spectrum underwent changes were obtained.

4. In contrast to the 2-component high-state spectrum described above, the seven time-average low-state spectra shown in Figure 2b (panels 12-18) are all consistent with a single power-law with photon indices of $\sim2.1-2.4$. A comparison of a typical low-state spectrum (Figure 2b panel 15) with the high-state spectrum is shown in Figure 3. In the soft $\gamma$-ray region above 200 keV, the spectrum of the low-intensity state is harder while that of the high-intensity is softer. The two spectra intersect at $\sim$400 keV.

5. Table 3 lists the best-fit parameters of two contrasting models, namely, a non-thermal power-law model and a thermal Comptonization model (Sunyaev $\&$ Titarchuk, 1980), to each of the 18 spectra shown in Figure 2a and 2b. The thermal Comptonization model fits the high-state spectra (panels 6-11) well, but not the power law. However, for the low-intensity state spectra (panels 12-18), while both the power-law and Compton models fit the spectra adequately, the power-law model is generally preferred because its model parameters are better constrained compared to those of the Compton models as indicated by the size of the errors.

6. The basic features of the high and low-intensity spectra described above were generally consistent with those observed by SIGMA (Revnivtsev et al. 1998) and OSSE (Grove et al. 1998). Namely, the high-intensity state spectra were better characterized by either the Compton model (Revnivtsev et al. 1998), or an Exponential Truncated Power Law (Grove et al. 1998), and the low-intensity spectra are better fitted with a power law. Tables 4 $\&$ 5a,b compare the best-fit model parameters measured by BATSE-EBOP with near contemporaneous results reported by SIGMA and OSSE, respectively. BATSE results generally agree well with those of SIGMA when contemporaneous measurements were directly compared. There was only one case when the comparison is not as good. This is during the period of SIGMA observations between TJD $9259-9274$ (29 September$-$14 October 1993) when BATSE collected no data during 11 days of this 16-day period from TJD $9264-9274$ due to CGRO reboost operations. BATSE and OSSE results also showed good agreement for one of the two periods when near contemporaneous measurements can be directly compared (e.g. TJD $9665-9671$; see Table 5a). However, for the 2nd spectrum that includes  OSSE observations A, B $\&$E (see Figure 1, panel 2), Grove et. al. (1998) stated that the spectrum measured during the 5th period between 1 and 14 February, 1995 (TJD $9749-9762$) was better fitted with the exponential truncated power law model than the power law model, similar to those measured in periods A $\&$ B.  Consequently, they combined the three spectra, and obtained the best-fit parameters for the combined spectrum (see Table 5a). BATSE results showed that neither model fits the combined spectrum well. While both a power-law and an exponential truncated power law fit the period E data adequately (see Table 5b), the best-fit parameters of these two models are consistent with a single power law with photon index of $\sim2.1-2.2$. On the other hand, the 4-day BATSE spectrum measured near contemporaneously with OSSE periods A $\&$ B showed a slightly better fit with the exponential truncated power law than the power law (Table 5b). 

\section{Discussion}

GRO~J$1719-24$, a low-mass x-ray (LMXB) binary system with a 14.7 hour periodicity,  consists of a $\sim$4.9 solar mass compact object and a $\sim$1.6 solar mass companion star (Della Valle, Mirabel $\&$ Rodriguez, 1994; Masett et al., 1996), is one of three galactic black-hole systems, along with GRO~J0422+32 (Ling $\&$ Wheaton 2003a) and Cygnus X-1 (Philips et al, 1996; Ling et al. 1997; McConnell et al. 2000, 2002) that displayed similar gamma-ray spectral characteristics when undergoing transitions between the high and low $\gamma$-ray intensity states (Ling $\&$ Wheaton 2004). 

$\bullet$ When these sources were in the high $\gamma$-ray intensity state ($\gamma_{2}$ for Cygnus X-1, Ling et al. 1987), their spectra showed two components: a Comptonized component from $\sim$30 keV to $\sim$200 keV followed by a soft power-law tail of photon index of $>$3 above $\sim$200 keV that extended to $\sim$500 keV for GRO~J$1719-24$, $\sim$1 MeV for Cygnus X-1 (Ling et al. 1997; McConnell et al. 2000),  and $\sim$600 keV for GRO~J0422+32 (Ling $\&$ Wheaton 2003a). For GRO~J$1719-24$, the 35-200 keV emission was likely to be produced by Compton scattering off hot electrons in the system with electron temperature $kT_e$ of $\sim$37 keV with optical depth $\tau$ of $\sim$2.9.

$\bullet$ When these sources were in the low $\gamma$-ray intensity state ($\gamma_{0}$ for Cygnus X-1, Ling et al. 1997), the Comptonized spectral shape below $\sim$200-300 keV vanished. The entire spectrum from 30 keV to $\sim$1.5 MeV was consistent with a power law with photon index of $\sim2.1-2.4$ for GRO~J$1719-24$. This is compared to $\sim1.8-2.1$ for GRO~J0422+32 (Ling $\&$ Wheaton 2003a), and $\sim$2.6-2.7 for Cygnus X-1 (Philips et al, 1996; Ling et al. 1997). Because the low-intensity spectrum is harder than the high-intensity spectrum above $\sim$200 keV, the two spectra intercept at $\sim$400 keV for GRO~J$1719-24$. This is compared to $\sim$600 keV for GRO~J0422+32 (Ling \& Wheaton 2003a) and $\sim$1 MeV for Cygnus X-1 (Ling $\&$ Wheaton 2004; McConnell et al. 2002).

$\bullet$ There is some weak evidence  ($\sim$2$\sigma$) that the spectrum changed from a power-law shape on TJD 9256 to  a braking ``thermal" shape on TJD 9257 during the rising phase of the 1993 transition. This suggests that if the system has undergone a change in its configuration, and the time scale for such a change could be on the order of one day. Because of the reboost operation, no similar information was obtained during the declining phase of the event $\sim$80 days later. 

$\bullet$ The high $\gamma$-ray intensity state ($\gamma_{2}$) for Cygnus X-1 was generally referred as the low/hard state, while the low $\gamma$-ray intensity state ($\gamma_{0}$) was referred to as the high/soft state, a description derived from the intensity and spectral shape of the 1-10 keV emission. We would like to point out that such labels are quite misleading in terms of gamma-ray emission. In the gamma-ray region between 30 keV and 1 MeV, the intensity is higher and the spectrum is softer above $\sim$200 keV for so-called low/hard x-ray state compared to the lower intensity and harder spectrum for the high/soft x-ray state. In order to prevent any unnecessary confusion, we have avoided using these standard labels that have been used in the past in describing x-ray spectrum for our discussion of the gamma-ray spectral features of these BH systems in this paper. We suggest that any future references in using these identifiers should specify the spectral region that they refer to.

$\bullet$ The two-component characteristic of the high-intensity state spectra, namely a thermal component below $\sim$200 keV and a non-thermal component above 200 keV,  shown in these three sources is different than the pure simple power-law spectrum shown in GRO~$J1655-40$ (Grove et al. 1998; Case et al. 2004), and a broken power law shown in GRS~1915+105 (Case et al. 2004), two other black-hole systems that are generally considered to be microquasars. These results confirm previous observations (Grove et al. 1998) that there are at least two classes of gamma-ray emitting BH candidates in our galaxy. 

The evidence for a persistent power-law component in both the high and low-intensity spectra of GRO~J$1719-24$ strongly suggest the presence of non-thermal processes in the system in both the high and low-intensity situations. The strong similarity of the gamma-ray characteristics shown in GRO~J$1719-24$ and Cygnus X-1 and GRO~J0422+32 raises the following key questions that need to be addressed. (1) What is the basic system configuration for interpreting the gamma-ray spectra? and (2) What are the physical processes responsible for the changes of the spectral shapes shown between the high and low-intensity spectra?

Over the past several years, a number of theories have been advanced in explaining the thermal and non-thermal emission observed in the high-intensity state spectra. Chakrabarti $\&$ Titarchuk (1995) and Turolla et al. (2002) have proposed that relativistic electrons with energies up to 1 MeV may be associated with free infall of matter onto the black hole in the converging flow region near the event horizon. The observed power-law spectral tail may possibly be produced by Compton up-scattering of photons produced in the system off these in-falling electrons. Meier (2001) suggested that non-thermal gamma-ray emission may be associated with jets, which he claimed is a natural consequence of accretion flows onto rotating black holes. More recently, Meier's (2004) model of Magnetically-Dominated Accretion Flows (MDAF) further develops the idea and integrates accretion and jet production as an extension of the ADAF (Esin et al. 1998) and Shakura $\&$ Sunyaev (1976) disk models. There were also suggestions for a hybrid thermal/non-thermal Comptonization model (Coppi 1998;  Gierlinski et al.1999). In this model, the electron distribution consisted of two components, a Maxwellian component with a temperature, $kT$, plus a non-thermal power-law component.  The acceleration of non-thermal electrons is independently taking place but is coupled to the background thermal plasma by Compton scattering and Coulomb collision processes. While such a model aimed to explain the two-component high-intensity gamma-ray spectra observed in BH binary systems such as GRO~J$1719-24$, GRO~J0422+32 and Cygnus X-1, no attempts have been made to explain how the spectrum evolved into a power law in the low-intensity scenario. 

Ling $\&$ Wheaton (2003a) suggested a possible system scenario for explaining both the high and low-intensity spectra. It is based on the ADAF model of Esin et al (1998) along with the source geometry envisioned by Poutanen $\&$ Coppi (1998) and others. Because the non-thermal power law was fully visible in the 35 keV to 1 MeV low-intensity spectrum, while only partially visible above 200 keV, and fully hidden behind the Comptonized spectrum below 200 keV in the high-intensity spectrum, they proposed a system, for the high-intensity scenario, that consists of a hot inner corona, a cooler outer thin disk, and a separate region that produces the power-law $\gamma$-ray emission.  We hereby refer to the latter the ``Non Thermal Emission Region (NTER)". NTER may include a jet (see Ling $\&$ Wheaton 2003a Figure 6; Ling $\&$ Wheaton 2004) and possibly also the converging flow region discussed above. Under such condition, the transition radius of the disk is $\sim$100 Schwarzschild radii from the black hole. Electrons in the hot corona up-scattered the low-energy photons produced both inside the corona as well as from the outer disk to form the Comptonized component that dominates the spectrum in the 35-200 keV range. These same electrons also down-scattered the high energy photons ($>$10 MeV) produced in NTER resulting in forming a softer power-law component observed in the 200 keV to 1 MeV range. Under the low-intensity scenario that could be triggered by a significant increase in the accretion rate, a large quantity of soft photon was produced in the disk that cooled and quenched the hot corona and moved the transition radius inward to a distance very close to the horizon. Under this condition, the Comptonized component below 200 keV disappeared and the entire 35-1000 keV spectrum is dominated by the unperturbed emission produced by NTER.

In summary, similar gamma-ray spectral characteristics and evolution between high and low-intensity transitions have now been observed in three galactic binary BH systems GRO~J$1719-24$, GRO~J0422+32 and Cygnus X-1. These results present new challenges and constraints to future theoretical work in this field.  We hope this work will stimulate a renewed interest to modelers to effectively address the questions posed by the new results presented in this paper.

\acknowledgements
	We wish to thank Gerald Fishman and his BATSE team for their support of the BATSE Earth Occultation investigation effort at JPL throughout the years, 
Dave Meier, Michael Cherry and Gary Case for their useful comments on this manuscript, and undergraduate students Robert Kern, 
Zachary Medin, and Juan Estrella for processing the data. The work 
described in this paper was carried out at the Jet Propulsion 
Laboratory, under the contract with the National Aeronautics and 
Space Administration.

\newpage
FIGURE CAPTIONS

\figcaption{The flux histories in the six broad-energy bands covering the period between 13 January 1993 (TJD 9000) and 10 October 1995 (TJD 10000) are shown. The resolutions are one day in the first two panels, two days in the 3rd panel, and five days in the last three panels. During the high-intensity period between TJD 9257 and TJD 9333, positive gamma-ray fluxes in each of the six energy bands were measured at $\sim$127, 108, 30, 9.9, 4.8, and 1.6$\sigma$ significance, respectively. Between TJD 9600 and 10000, the source flared again five different times (marked as ``14", ``15", ``16", ``17" and ``18") and the 35-100 keV fluxes were measured at a lower level of $\sim$354 mCrab, $\sim$354 mCrab, $\sim$197 mCrab, $\sim$189 mCrab, and $\sim$290 mCrab respectively. Data gaps shown between TJD 9264 and 9341 were periods when the spacecraft underwent reboost operations. Shown also in the 2nd panel the markers for the six observational periods of SIGMA (label ``a"-``f"), and the five observational periods of OSSE (label ``A"-``E")}

\figcaption{(a) Nine daily spectra measured from TJD 9249 to 9260 during the rising phase of the source transition from the quiescent state (Q1) to the high state. In each panel, the solid line is the best-fit model to the spectral data measured by the ``source-viewing" LADs. Pertinent information of the model fit is also displayed in each panel and listed in Table 3. The spectrum changed from a power-law shape prior to TJD 9256 (panels 1-5) to a Comptonized shape below 200 keV on TJD 9257 (panel 6) and stayed in this shape for the rest of the rising phase (panels 7-9). (b) The two averaged high-state spectra integrated over TJD $9262-9299$ and TJD $9301-9333$ (panels 10 and 11), respectively, can be described as having two components: a Comptonized shape below $\sim$200 keV and a soft power-law tail above $\sim$200 keV that extends to $\sim400-500$ keV. The seven low-state spectra measured in $1994-1995$ on the other hand have a power-law shape with photon indices of $\sim2.1-2.4$.}

\figcaption{A comparison of a typical low-state (Figure 2b panel 15) spectrum with the average of the two high-state spectra shown in Figure 2b panels 10-11. The high and low-state spectra intersect at $\sim$400 keV, compared to those observed for GRO~J0422+32 at $\sim$600 keV (Ling and Wheaton, 2003), and $\sim$1 MeV for Cygnus X-1 (Ling $\&$ Wheaton 2004; McConnell et al. 2002).}

\setcounter{figure}{0}
\nonumber
\begin{figure}[ht]
\centering
\includegraphics[scale=0.850]{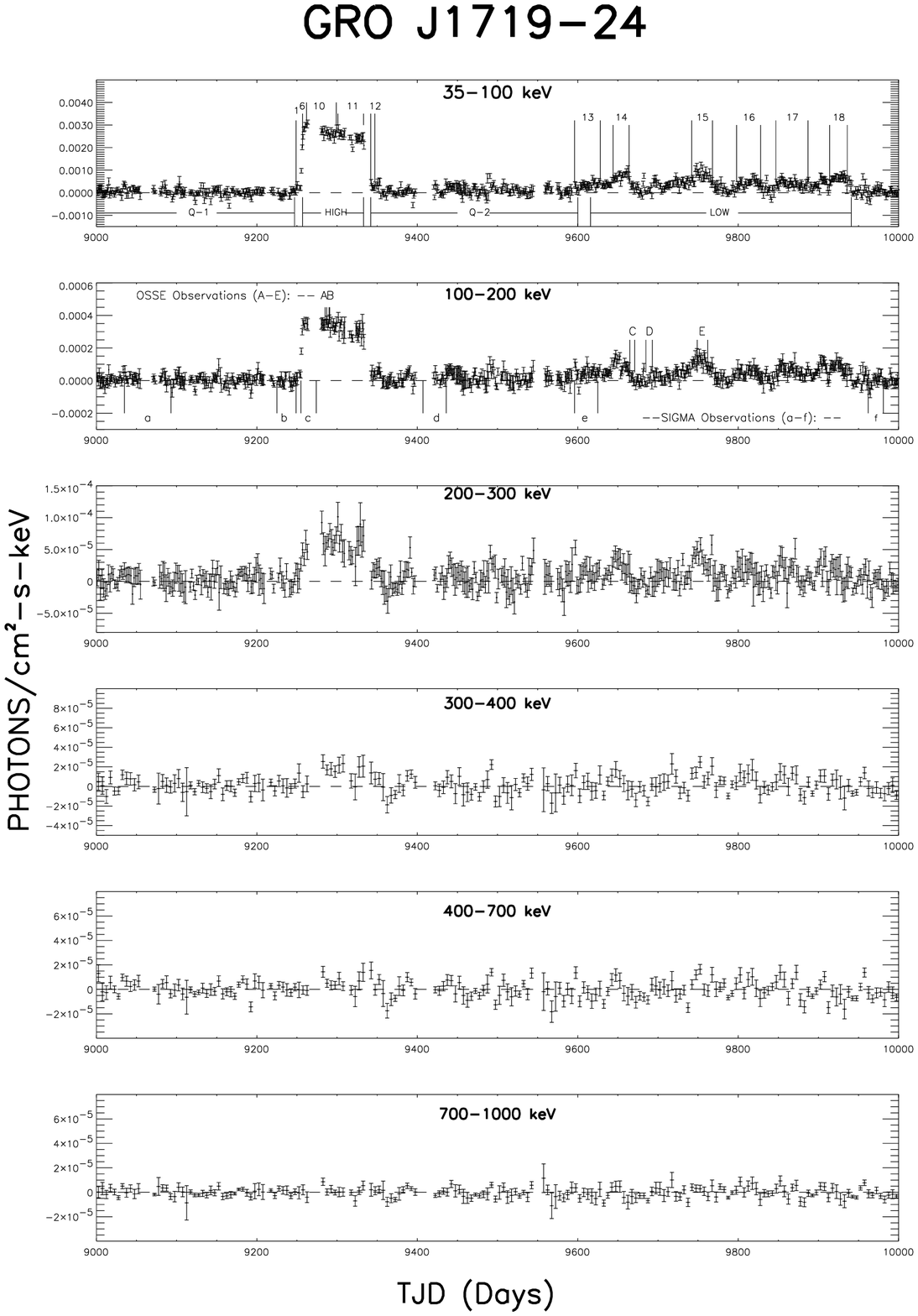}
\caption{Figure 1}
\end{figure}
\begin{figure}[ht]
\centering
\includegraphics[scale=0.850]{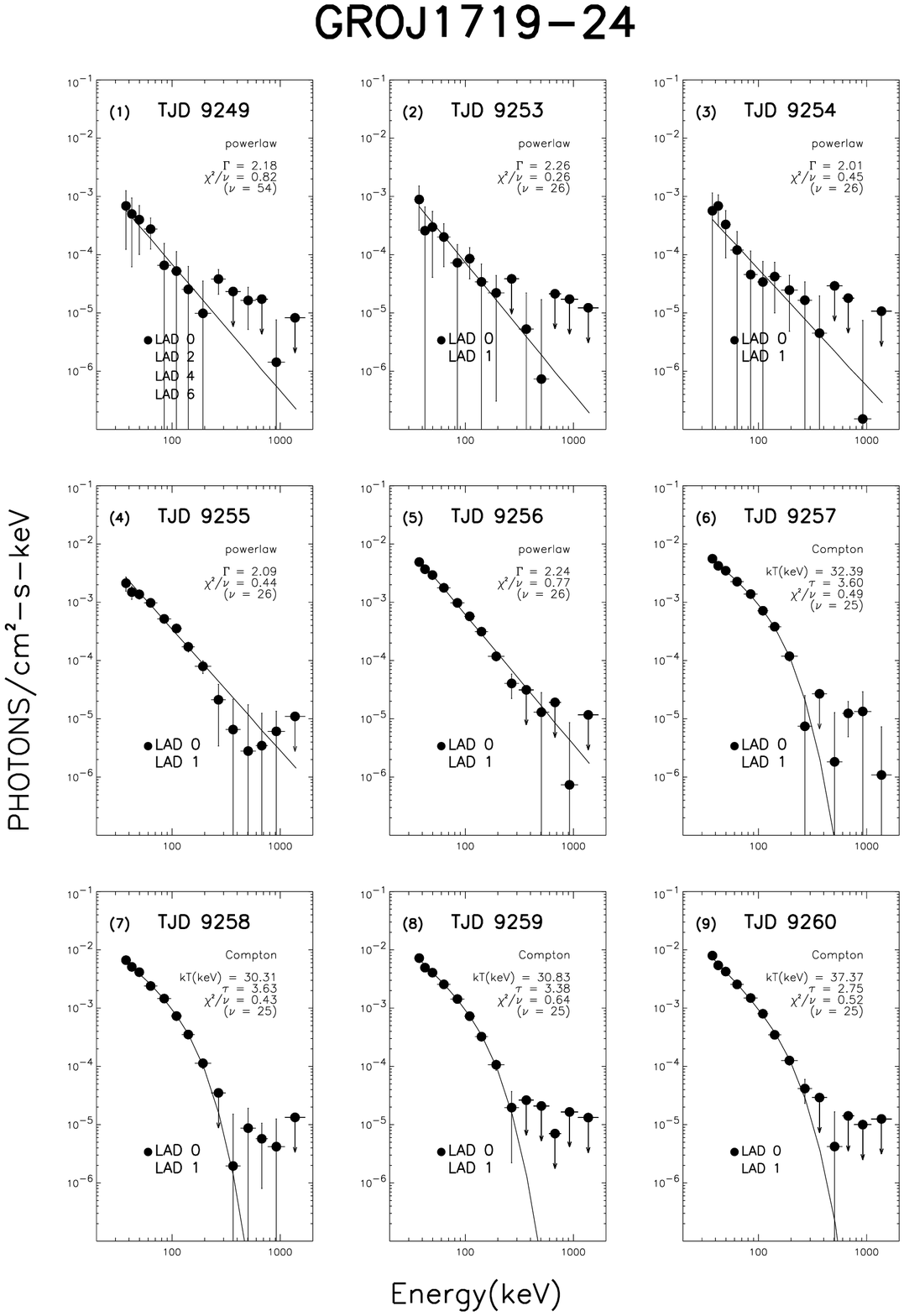}
\caption{Figure 2a}
\end{figure}
\begin{figure}[ht]
\centering
\includegraphics[scale=0.850]{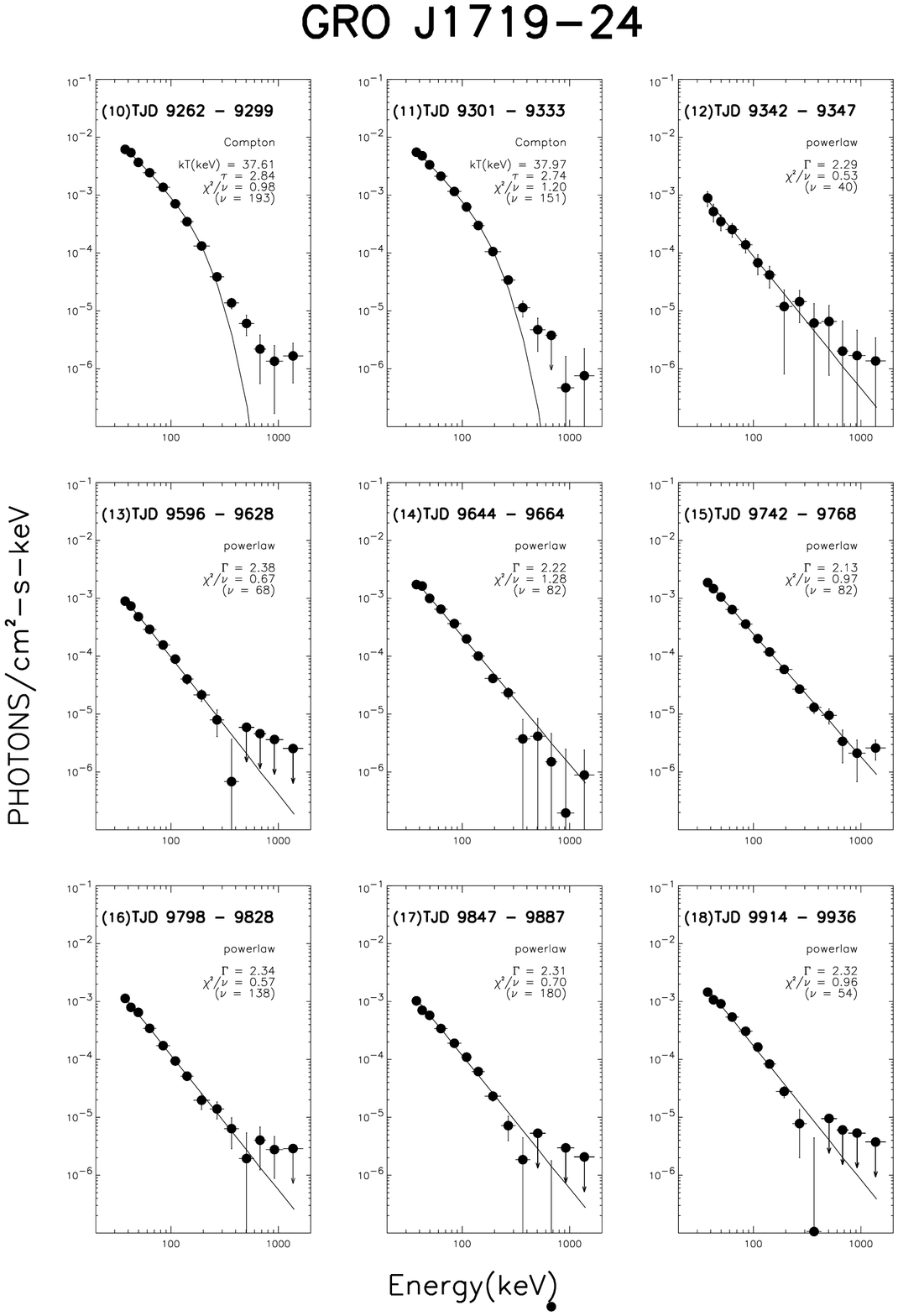}
\caption{Figure 2b}
\end{figure}
\begin{figure}[ht]
\centering
\includegraphics[scale=0.850]{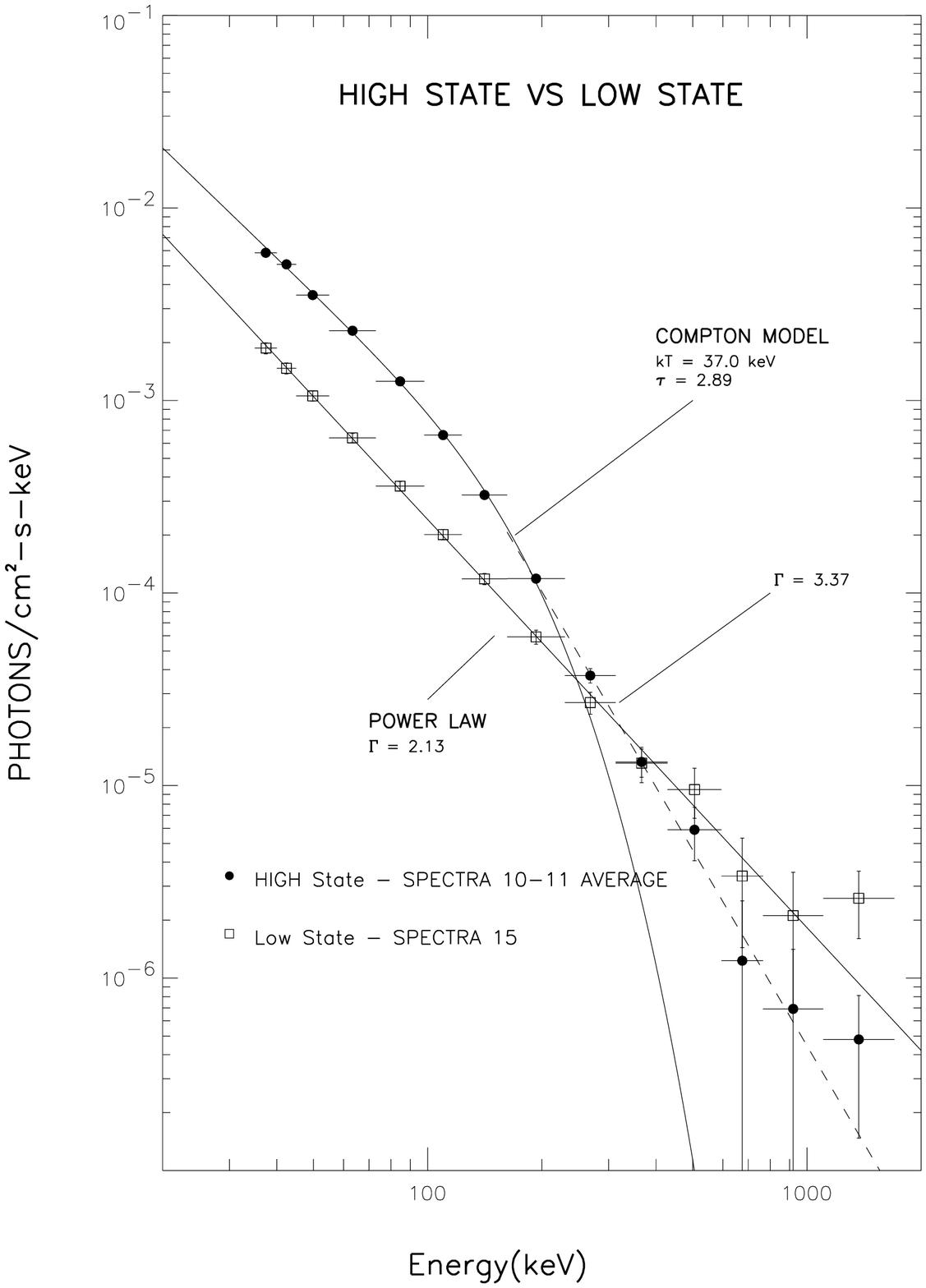}
\caption{Figure 3}
\end{figure}
\clearpage

\begin{deluxetable}{c c c c c} %
\tabletypesize{\scriptsize}%

\tablewidth{0pt}%
\tablecaption{Average gamma-ray fluxes for the quiescent and high Intensity spectral states}%
\tablenum{1}%

\tablehead{ &  & \colhead{Quiescent Period 1 (Q-1)} & \colhead{High-State Period} & \colhead{Quiescent Period 2 (Q-2)} \\

\colhead{Energy} & \colhead{Crab Flux} & \colhead{(TJD 9000-9247)} & \colhead{(TJD 9262-9333)} & \colhead{(TJD 9342-9600)} \\

\colhead{(keV)} & \colhead{(Reference 1)} & \colhead{(186-day Integration)} & \colhead{(42-day Integration)} & \colhead{(182-day Integration)} }%

\startdata
  35-100  &  $(208.6\pm0.3)  \times 10^{-5}$  &  $(2.19 \pm0.74) \times 10^{-5}$  &  $(257.34\pm2.04) \times 10^{-5}$  &  $(9.28 \pm0.88) \times 10^{-5}$  \\%
 100-200  &  $(30.66\pm0.06) \times 10^{-5}$  &  $(1.92 \pm1.51) \times 10^{-6}$  &  $(32.30 \pm0.46) \times 10^{-5}$  &  $(1.43 \pm0.19) \times 10^{-5}$  \\%
 200-300  &  $(9.20 \pm0.03) \times 10^{-5}$  &  $(1.66 \pm0.96) \times 10^{-6}$  &  $(5.98  \pm0.30) \times 10^{-5}$  &  $(3.30 \pm1.18) \times 10^{-6}$  \\%
 300-400  &  $(3.71 \pm0.03) \times 10^{-5}$  &  $(5.61 \pm7.62) \times 10^{-7}$  &  $(1.65  \pm0.23) \times 10^{-5}$  &  $(9.68 \pm9.44) \times 10^{-7}$  \\%
 400-700  &  $(14.88\pm0.24) \times 10^{-6}$  &  $(1.60 \pm4.89) \times 10^{-7}$  &  $(4.12  \pm1.29) \times 10^{-6}$  &  $(1.05 \pm6.05) \times 10^{-7}$  \\%
700-1000  &  $(5.44 \pm0.17) \times 10^{-6}$  &  $(-1.65\pm3.24) \times 10^{-7}$  &  $(1.00  \pm0.89) \times 10^{-6}$  & $(-5.40\pm4.09) \times 10^{-7}$  \\%
\enddata

\tablecomments{All flux values are given in units of
photon~cm$^{-2}$~s$^{-1}$~keV$^{-1}$}

\tablerefs{(1) Ling and Wheaton 2003b.}

\end{deluxetable}


\begin{deluxetable}{c c c c c c} %
\tabletypesize{\scriptsize}%

\tablewidth{0pt}%
\tablecaption{Average gamma-ray fluxes for the five low-intensity spectral states}%
\tablenum{2}%

\tablehead{ & \colhead{Low-State Period-14} & \colhead{Low-State Period-15} & \colhead{Low-State Period-16} & \colhead{Low-State Period-17} & \colhead{Low-State Period-18} \\

\colhead{Energy} & \colhead{(TJD 9644-9664)} & \colhead{(TJD 9742-9768)} & \colhead{(TJD 9798-9828)} & \colhead{(TJD 9847-9887)} & \colhead{(TJD 9914-9936)} \\

\colhead{(keV)} & \colhead{(19-day Integration)} & \colhead{(26-day Integration)} & \colhead{(28-day Integration)} & \colhead{(38-day Integration)} & \colhead{(22-day Integration)} }%

\startdata
  35-100  &  $(73.83\pm2.65) \times 10^{-5}$ &  $(73.96\pm2.63) \times 10^{-5}$  &  $(41.22\pm2.55) \times 10^{-5}$  &  $(39.46\pm2.13) \times 10^{-5}$  &  $(60.53\pm3.06) \times 10^{-5}$ \\%
 100-200  &  $(9.97 \pm0.61) \times 10^{-5}$ &  $(11.28\pm0.51) \times 10^{-5}$  &  $(4.92 \pm0.55) \times 10^{-5}$  &  $(5.88 \pm0.46) \times 10^{-5}$  &  $(7.99 \pm0.66) \times 10^{-5}$ \\%
 200-300  &  $(2.86 \pm0.45) \times 10^{-5}$ &  $(3.61 \pm0.31) \times 10^{-5}$  &  $(1.52 \pm0.42) \times 10^{-5}$  &  $(1.20 \pm0.29) \times 10^{-5}$  &  $(1.29 \pm0.50) \times 10^{-5}$ \\%
 300-400  &  $(0.77 \pm0.41) \times 10^{-5}$ &  $(1.49 \pm0.26) \times 10^{-5}$  &  $(0.80 \pm0.35) \times 10^{-5}$  &  $(0.31 \pm0.26) \times 10^{-5}$  &  $(0.17 \pm0.42) \times 10^{-5}$ \\%
 400-700  &  $(0.43 \pm0.29) \times 10^{-5}$ &  $(0.81 \pm0.18) \times 10^{-5}$  &  $(0.39 \pm0.24) \times 10^{-5}$  &  $(0.03 \pm0.18) \times 10^{-5}$  &  $(-0.53\pm0.31) \times 10^{-5}$ \\%
700-1000  &  $(0.10 \pm0.23) \times 10^{-5}$ &  $(0.25 \pm0.13) \times 10^{-5}$  &  $(0.34 \pm0.17) \times 10^{-5}$  &  $(-0.04\pm0.13) \times 10^{-5}$  &  $(-0.13\pm0.22) \times 10^{-5}$ \\%
\enddata

\tablecomments{All flux values are given in units of
photon~cm$^{-2}$~s$^{-1}$~keV$^{-1}$}

\end{deluxetable}

\hspace{-1.0in}
\begin{deluxetable}{c@{}c@{}c c@{}c@{}c@{}c@{}c@{}c@{}c@{}c@{}c@{}c@{}c c@{}c@{}c} %
\rotate%
\tabletypesize{\tiny}%

\tablewidth{0pt}%
\tablecaption{Best-fit Model Parameters}%
\tablenum{3}%

\tablecolumns{17}%
\tablehead{%
&            &            &            & & & &  &  &  &  &  &  &  &  \\ %
&            &            &            & & & &  & \multicolumn{3}{c}{\vspace{-8pt}Power Law Model ($AE^{-\Gamma}$)} & \multicolumn{4}{c}{Compton Model (ST)}  \\ %
           &            &            &            & & & &  & \multicolumn{3}{c}{\rule{1.6in}{0.5pt}}              & \multicolumn{4}{c}{\rule{2.2in}{0.5pt}} \\ %

\colhead{Spectral}                   & \colhead{Spectral}                   &                                 &                                & \colhead{Days~of}                       &                                 & \colhead{35--100 keV}            & \colhead{Hardness Flux Ratios}            &                                      &                                      &                                          & \colhead{kT}                      &                                    &                                      &                                     \\ %
\colhead{\raisebox{1.0ex}{Number}}   & \colhead{\raisebox{1.0ex}{State\tablenotemark{a}}}    & \colhead{\raisebox{1.0ex}{TJD}} & \colhead{\raisebox{1.0ex}{VP}} & \colhead{\raisebox{1.0ex}{Integration}} & \colhead{\raisebox{1.0ex}{LAD}} & \colhead{\raisebox{1.0ex}{Flux}} & \colhead{\raisebox{1.0ex}{(150-300)/(35-150) keV)}} & \colhead{\raisebox{1.0ex}{$\Gamma$}} & \colhead{\raisebox{1.0ex}{DOF($v$)}} & \colhead{\raisebox{1.0ex}{$\chi^{2}/v$}} & \colhead{\raisebox{1.0ex}{(keV)}} & \colhead{\raisebox{1.0ex}{$\tau$}} & \colhead{\raisebox{1.5ex}{DOF($v$)}} & \colhead{\raisebox{1.0ex}{$\chi^{2}/v$}} }%

\startdata
1   &  RP  &  9249              &  302.3    &   1  &  0,2,4,6  &   2.38$\pm$1.19  &  $(1.34\pm1.21)\times10^{-1}$   &  2.18$\pm$0.48   &   26  &  0.82  &  \nodata          &  \nodata        &  \nodata  &  \nodata  \\%
2   &  RP  &  9253              &  303.2    &   1  &  0,1      &   2.17$\pm$1.11  &  $(0.69\pm1.03)\times10^{-1}$   &  2.26$\pm$0.63   &   26  &  0.26  &  69.5$\pm$170.2   &  1.74$\pm$4.64  &   25      &  0.27     \\%
3   &  RP  &  9254              &  303.2    &   1  &  0,1      &   2.06$\pm$1.04  &  $(1.48\pm1.20)\times10^{-1}$   &  2.01$\pm$0.72   &   26  &  0.45  &  81.0$\pm$183.0   &  2.25$\pm$4.76  &   25      &  0.44     \\%
4   &  RP  &  9255              &  303.2    &   1  &  0,1      &   9.71$\pm$1.03  &  $(8.68\pm2.16)\times10^{-2}$   &  2.09$\pm$0.12   &   26  &  0.44  &  43.0$\pm$12.5    &  3.15$\pm$0.98  &   25      &  0.2      \\%
5   &  RP  &  9256              &  303.2    &   1  &  0,1      &  20.12$\pm$1.04  &  $(7.21\pm1.08)\times10^{-2}$   &  2.24$\pm$0.07   &   26  &  0.77  &  43.8$\pm$7.9     &  2.70$\pm$0.53  &   25      &  0.15     \\%
6   &  RP  &  9257              &  303.2    &   1  &  0,1      &  24.75$\pm$1.01  &  $(5.07\pm0.86)\times10^{-2}$   &  2.32$\pm$0.06   &   26  &  2.29  &  32.4$\pm$3.8     &  3.60$\pm$0.53  &   25      &  0.49     \\%
7   &  RP  &  9258              &  303.2    &   1  &  0,1      &  28.20$\pm$1.04  &  $(4.09\pm0.78)\times10^{-2}$   &  2.39$\pm$0.06   &   26  &  2.15  &  30.3$\pm$3.3     &  3.63$\pm$0.55  &   25      &  0.43     \\%
8   &  RP  &  9259              &  303.2    &   1  &  0,1      &  28.45$\pm$1.03  &  $(4.36\pm0.77)\times10^{-2}$   &  2.46$\pm$0.06   &   26  &  2.36  &  30.8$\pm$3.5     &  3.38$\pm$0.50  &   25      &  0.64     \\%
9   &  RP  &  9260              &  303.2    &   1  &  0,1      &  30.04$\pm$1.03  &  $(5.27\pm0.75)\times10^{-2}$   &  2.40$\pm$0.05   &   26  &  1.78  &  37.4$\pm$5.4     &  2.75$\pm$0.39  &   25      &  0.52     \\%
10  &  HS  &  9262--9299        &           &  20  &           &  27.15$\pm$0.28  &  $(5.97\pm0.23)\times10^{-2}$   &  2.38$\pm$0.01   &  194  &  2.61  &  37.6$\pm$1.4     &  2.84$\pm$0.12  &  193      &  0.98     \\%
    &      &  9262--9263        &  303.4    &      &  0,4      &                  &                                 &                  &       &        &                   &                 &           &           \\%
    &      &  9280--9284        &  304      &      &  0,1,2,3  &                  &                                 &                  &       &        &                   &                 &           &           \\%
    &      &  9286--9292        &  305      &      &  0,1,2,3  &                  &                                 &                  &       &        &                   &                 &           &           \\%
    &      &  9294--9299        &  306      &      &  0,1,2,3  &                  &                                 &                  &       &        &                   &                 &           &           \\%
11  &  HS  &  9301--9333        &           &  26  &           &  24.14$\pm$0.26  &  $(5.55\pm0.27)\times10^{-2}$   &  2.38$\pm$0.02   &  152  &  2.62  &  38.0$\pm$1.7     &  2.74$\pm$0.13  &  151      &  1.2      \\%
    &      &  9301--9306        &  307      &      &  0,1,2,3  &                  &                                 &                  &                  &       &        &                   &                 &           &           \\%
    &      &  9308--9309        &  308      &      &  0,1      &                  &                                 &                  &                  &       &        &                   &                 &           &           \\%
    &      &  9317-9320         &  308.6    &      &  0,1      &                  &                                 &                  &       &        &                   &                 &           &           \\%
    &      &  9323--9333        &  310      &      &  1,3,7    &                  &                                 &                  &       &        &                   &                 &           &           \\%
12  &  LS  &  9342--9347        &  312      &   6  &  0,1,2    &  2.84$\pm$0.50   &  $(8.21\pm4.00)\times10^{-2}$   &  2.29$\pm$0.24   &   40  &  0.53  &  170.4$\pm$639.9  &   0.70$\pm$2.82  &  151      &  0.56     \\%
13  &  LS  &  9596--9628        &           &  22  &           &  3.48$\pm$0.21   &  $(7.42\pm1.51)\times10^{-2}$   &  2.38$\pm$0.08   &   68  &  0.67  &  61.4$\pm$26.1    &  1.75$\pm$0.68  &  151      &  0.6      \\%
    &      &  9596--9614        &  338.5    &      &  2,3,7    &                  &                                 &                  &       &        &                   &                 &           &           \\%
    &      &  9616--9628        &  339      &      &  0,4      &                  &                                 &                  &       &        &                   &                 &           &           \\%
14  &  LS  &  9644--9664        &           &      &           &  7.42$\pm$0.27   &  $(7.81\pm0.95)\times10^{-2}$   &  2.22$\pm$0.05   &   82  &  1.28  &  57.6$\pm$12.1    &  2.00$\pm$0.39  &  151      &  1.2      \\%
    &      &  9644--9649        &  402      &      &  2,6      &                  &                                 &                  &       &        &                   &                 &           &           \\%
    &      &  9651--9656        &  402.5    &      &  0,2      &                  &                                 &                  &       &        &                   &                 &           &           \\%
    &      &  9658--9664        &  403      &      &  0,1      &                  &                                 &                  &       &        &                   &                 &           &           \\%
15  &  LS  &  9742--9768        &           &  26  &           &  7.49$\pm$0.27   &  $(9.86\pm0.71)\times10^{-2}$   &  2.13$\pm$0.048  &   82  &  0.97  &  125.1$\pm$33.9   &  1.11$\pm$0.32  &  151      &  1.06     \\%
    &      &  9742--9761        &  410      &      &  0,1,2,3  &                  &                                 &                  &       &        &                   &                 &           &           \\%
    &      &  9763--9768        &  411.1    &      &  3,7      &                  &                                 &                  &       &        &                   &                 &           &           \\%
16  &  LS  &  9798--9828        &           &  28  &           &  4.17$\pm$0.26   &  $(6.74\pm1.57)\times10^{-2}$   &  2.34$\pm$.09    &  138  &  0.57  &  103.9$\pm$78.2   &  1.08$\pm$0.86  &  151      &  0.59     \\%
    &      &  9798--9804        &  414      &      &  0,4,5    &                  &                                 &                  &       &        &                   &                 &           &           \\%
    &      &  9806--9810        &  414.3    &      &  0,2,4    &                  &                                 &                  &       &        &                   &                 &           &           \\%
    &      &  9812--9817        &  419.1    &      &  1,5      &                  &                                 &                  &       &        &                   &                 &           &           \\%
    &      &  9819--9828        &  415      &      &  4,5      &                  &                                 &                  &       &        &                   &                 &           &           \\%
17  &  LS  &  9847--9887        &           &  38  &           &  3.96$\pm$022    &  $(7.12\pm1.14)\times10^{-2}$   &  2.31$\pm$.07    &  180  &  0.7   &  51.2$\pm$11.9    &  2.24$\pm$0.51  &  151      &  0.63     \\%
    &      &  9847--9859        &  419      &      &  5,7      &                  &                                 &                  &       &        &                   &                 &           &           \\%
    &      &  9861--9873        &  420      &      &  1,5,7    &                  &                                 &                  &       &        &                   &                 &           &           \\%
    &      &  9875--9880        &  421      &      &  0,2,4,6  &                  &                                 &                  &       &        &                   &                 &           &           \\%
    &      &  9882--9887        &  422      &      &  0,2,4,6  &                  &                                 &                  &       &        &                   &                 &           &           \\%
18  &  LS  &  9914--9936        &           &  22  &           &  6.06$\pm$0.31   &  $(5.47\pm1.16)\times10^{-2}$   &  2.32$\pm$.06    &   54  &  0.96  &  37.4$\pm$6.3     &  2.95$\pm$0.55  &  151      &  0.58     \\%
    &      &  9914--9922        &  424      &      &  4,6      &                  &                                 &                  &       &        &                   &                 &           &           \\%
    &      &  9924--9936        &  425      &      &  5,7      &                  &                                 &                  &       &        &                   &                 &           &           \\%
\enddata

\tablenotetext{a}{RP = Rising Phase; HS = High State; LS = Low
State}

\end{deluxetable}


\begin{deluxetable}{c c c c c c c c} %
\tabletypesize{\scriptsize}%

\tablewidth{0pt}%
\tablecaption{Comparison of best-fit model parameters of contemporaneous measurements between BATSE and SIGMA.}%
\tablenum{4}%

\tablecolumns{17}%
\tablehead{%
           &            &            & \multicolumn{2}{c}{\vspace{-8pt}Power Law ($\sim E^{-\Gamma}$)} & \multicolumn{3}{c}{Compton Model}       \\ %
           &            &            & \multicolumn{2}{c}{\rule{1.3in}{0.5pt}}                         & \multicolumn{3}{c}{\rule{2.5in}{0.5pt}} \\ %

           &                                      &                                 &                                      &                                             & \colhead{kT}                      &                                    &                                           \\ %
           & \colhead{\raisebox{1.0ex}{Dates}}    & \colhead{\raisebox{1.0ex}{TJD}} & \colhead{\raisebox{1.0ex}{$\Gamma$}} & \colhead{\raisebox{1.0ex}{$\chi^{2}$(DOF)}} & \colhead{\raisebox{1.0ex}{(keV)}} & \colhead{\raisebox{1.0ex}{$\tau$}} & \colhead{\raisebox{1.0ex}{$\chi^{2}$(DOF)}} }%

\startdata %
SIGMA  &  Sep 25--26, 1993        &  9255--9256  &  2.00$\pm$0.07  &  76.0(57)    &  55 ($+33$,$-10$)      &  2.4 ($+0.6$,$-0.7$)  &  73(56)    \\%
BATSE  &  Sep 25--26, 1993        &  9255--9256  &  2.19$\pm$0.06  &  106.8(54)   &  45.7$\pm$8.3          &  2.69$\pm$0.46        &  85.7(53)  \\%
&&&&&&&\\%
SIGMA  &  Sep 26--28, 1993        &  9256--9258  &  2.33$\pm$0.05  &  88.2(57)    &  37 ($+6$,$-4$)        &  2.7$\pm$04           &  66(56)    \\%
BATSE  &  Sep 26--28, 1993        &  9256--9258  &  2.33$\pm$0.03  &  172.3(82)   &  33.9$\pm$2.4          &  3.37$\pm$0.31        &  71.5(81)  \\%
&&&&&&&\\%
SIGMA  &  Sep 29 -- Oct 14, 1993  &  9259--9274  &  2.34$\pm$0.01  &  430.8(57)   &  42.6 ($+1.4$,$-1.3$)  &  2.3$\pm$0.1          &  90(56)    \\%
BATSE  &  Sep 29--30, 1993 $\&$        &  9259--9260 $\&$ &  2.44$\pm$0.03  &  212.8(109)  &  32.6$\pm$2.1          &  3.15$\pm$0.23        &  64.1(109) \\%
       &  Oct 2--3, 1993          &  9262--9263  &                 &              &                        &                       &            \\%
&&&&&&&\\%
SIGMA  &  Sep 1--30, 1994         &  9596--9625  &  2.19$\pm$0.15  &  60(57)      &  68 ($+\infty$,$-28$)  &  1.7$\pm$1.2          &  59(56)    \\%
BATSE  &  Sep 1--19, 1994 $\&$         &  9596-9614 $\&$  &  2.38$\pm$0.08  &  45.5(68)    &  61.4$\pm$26.1         &  1.75$\pm$0.68        &  40.3(68)  \\%
       &  Sep 21--30, 1994        &  9616--9625  &                 &              &                        &                       &            \\%
\enddata

\end{deluxetable}


\begin{deluxetable}{c c c c c c c c} %
\tabletypesize{\scriptsize}%

\tablewidth{0pt}%
\tablecaption{Comparison of best-fit model parameters of contemporaneous measurements between BATSE and OSSE.}%
\tablenum{5a}%

\tablecolumns{17}%
\tablehead{%
           &            &            & \multicolumn{2}{c}{\vspace{-8pt}Power Law ($\sim E^{-\Gamma}$)} & \multicolumn{3}{c}{Exponential Trucated Power Law ($\sim E^{-\Gamma}e^{-E/E_{f}}$)}       \\ %
           &            &            & \multicolumn{2}{c}{\rule{1.3in}{0.5pt}}                         & \multicolumn{3}{c}{\rule{2.7in}{0.5pt}} \\ %

           &                                      &                                 &                                      &                                             &                                      & \colhead{\raisebox{1.0ex}{$E_{f}$}} &                                           \\ %
           & \colhead{\raisebox{1.0ex}{Dates}}    & \colhead{\raisebox{1.0ex}{TJD}} & \colhead{\raisebox{1.0ex}{$\Gamma$}} & \colhead{\raisebox{1.0ex}{$\chi^{2}$(DOF)}} & \colhead{\raisebox{1.0ex}{$\Gamma$}} & \colhead{\raisebox{1.0ex}{(keV)}} & \colhead{\raisebox{1.0ex}{$\chi^{2}$(DOF)}} }%

\startdata %
OSSE-A &  Oct 25-27, 1993 $\&$  &  9285-9287 $\&$ &                 &                 &  1.53$\pm$0.06   &  115$\pm$8          &  not available  \\%
    -B &  Oct 30-31, 1993 $\&$ &  9290-9291 $\&$ &                 &                 &                  &                     &                 \\%
    -E &  Feb 1-14, 1995   &  9749-9762  &                 &                 &                  &                     &                 \\%
BATSE-A &  Oct 26-27, 1993 $\&$ &  9286-9287 $\&$ &  2.26$\pm$0.02  &  1007(110)      &  1.71$\pm$0.13   &  197.2$\pm$47.4     &  985(109)       \\%
     -B &  Oct 30-31, 1993 $\&$ &  9290-9291 $\&$ &                 &                 &                  &                     &                 \\%
-E &  Feb 1-13, 1995   &  9749-9762  &                 &                 &                  &                     &                 \\%
& & & & & & & \\%
OSSE   &  Nov 9-15, 1994   &  9665-9671  &  2.42$\pm$0.08  &  not available  &                  &                     &                 \\%
BATSE  &  Nov 10-15, 1994  &  9666-9670  &  2.51$\pm$0.10  &  33.5(26)       &                  &                     &                 \\%
\enddata

\end{deluxetable}
\begin{deluxetable}{c c c c c c c c} %
\tabletypesize{\scriptsize}%

\tablewidth{0pt}%
\tablecaption{Best-fit model parameters of the BATSE spectrum measured during the three OSSE observational periods A, B $\&$E.}%
\tablenum{5b}%

\tablecolumns{17}%
\tablehead{%
           &            &            & \multicolumn{2}{c}{\vspace{-8pt}Power Law ($\sim E^{-\Gamma}$)} & \multicolumn{3}{c}{Exponential Trucated Power Law ($\sim E^{-\Gamma}e^{-E/E_{f}}$)}       \\ %
           &            &            & \multicolumn{2}{c}{\rule{1.3in}{0.5pt}}                         & \multicolumn{3}{c}{\rule{2.7in}{0.5pt}} \\ %

           &                                      &                                 &                                      &                                             &                                      & \colhead{\raisebox{1.0ex}{$E_{f}$}} &                                           \\ %
           & \colhead{\raisebox{1.0ex}{Dates}}    & \colhead{\raisebox{1.0ex}{TJD}} & \colhead{\raisebox{1.0ex}{$\Gamma$}} & \colhead{\raisebox{1.0ex}{$\chi^{2}$(DOF)}} & \colhead{\raisebox{1.0ex}{$\Gamma$}} & \colhead{\raisebox{1.0ex}{(keV)}} & \colhead{\raisebox{1.0ex}{$\chi^{2}$(DOF)}} }%

\startdata %
BATSE-A&  Oct 26-27, 1993 $\&$ &  9286-9287 $\&$ &  2.36$\pm$0.03  &  238.2(222)     &  1.24$\pm$0.16   &  88.0$\pm$14.5      &  164.5(221)     \\%
BATSE-B&  Oct 30-31, 1993  &  9290-9291  &                 &                 &                  &                     &                 \\%
& & & & & & & \\%
BATSE-E&  Feb 1-13, 1995   &  9749-9761  &  2.19$\pm$0.08  &  35.4(54)       &  2.11$\pm$0.131  &  1472.7$\pm$2546.0  &  35.1(53)       \\%
& & & & & & & \\%
\enddata

\end{deluxetable}
\clearpage


\begin{references}
\reference{Arnaud 1996}Arnaud, K. A., 1996, in ASP Conf. Proc. 101, Astronomical Data Analysis Software and Systems V, ed. G. Jacoby $\&$ J. Barnes (San Francisco; ASP), 17.
\reference{Ballet et al. 1993}Ballet, J., Denis, M., Gilfanov, M., $\&$ Sunyaev, R. 1993, IAU Circ. 5874
\reference{Borozdin et al. 1995}Borozdin, K., Alexandrovich, N., $\&$ Sunyaev, R., 1995, IAU Circ. 6141.
\reference{Case et al. 2004} Case, G. L., Cherry, M. L., Fannin, C., Rodi, J.,  Ling, J. C., $\&$ Wheaton, Wm, A.,  2004, 5th Microquasars Conference Proceedings, Beijing, China June 7-13 \& astro-ph/0409306 
\reference{Chakrabarti $\&$ Titarchuk 1995}Chakrabarti, S. K., $\&$ Titarchuk, L., 1995, ApJ, 455, 623.
\reference{Churazov et al. 1994}Churazov, E., Gilfanov, M., Ballet, J., $\&$ Jourdain, E. 1994, IAU Circ., No. 6083.
\reference{Coppi 1998}Coppi, P. S. , 1998, ``The Physics of Hybrid Thermal/Non-thermal Plasmas" in High energy Porcesses in Accreting Black Holes, eds. J Poutanen and R. Svensson, ASP Conf. Series, Vol. 161, p. 375 (astro-ph/9903158)
\reference{Della Valle et al. 1994}Della Valle, M., Mirabel, F., $\&$ Rodriguez, L. F., 1994, A$\&$A, 290, 803.
\reference{Esin et al. 1998}Esin, A. A., Narayan, R., Cui, W., Grove, J. E., $\&$ Zhang, S., 1998, ApJ, 505, 854.
\reference{Gierlinski et al.1999}Gierlinski, M., Zdziarski, A. A., Poutanen, J., Coppi, P. S. Ebisawa, K., and Johnson, W. N. 1999, MNRAS, 309, 496.
\reference{Grove et al.1998}Grove, J. E., Johnson, W. N., Kroeger, R. A., McNaron-Brown, K., $\&$ Skibo, J. G., 1998, ApJ, 500, 899.
\reference{Harmon et al. 1993a} Harmon, B. A., Zhang, S. N., Paciesas, W. S., \& Fishman, G. J., 1993a, IAU Circ. 5874
\reference{Harmon et al. 1993b} Harmon, B. A., Fishman, G. J., Paciesas, W. S., \& Zhang, S. N., 1993b, IAU Circ. 5900 
\reference{Hjellming et al. 1996}Hjellming R. M., Rupen, M. P., Shrader, C. R., Campbell-Wilson, D., Hunstead, R. W., $\&$ McKay, D. J. 1996, ApJ, 471, L105.
\reference{Ling & Wheaton, 2003a}Ling, J. C., and Wheaton, W. A., 2003a, ApJ,  384, 399.
\reference{Ling & Wheaton, 2003b}Ling, J. C., and Wheaton, W. A., 2003b, ApJ,  598, 334.
\reference{Ling & Wheaton, 2004}Ling, J. C., and Wheaton, W. A., 2004, 5th Microquasars Conference Proceedings, Beijing, China, June 7-13, 2004, \& astro-ph/0409307 
\reference{Ling et al. 1996}Ling, J. C., Wheaton, W. A., Mahoney, W. A., Skelton, R. T., Radocinski, R. G., and Wallyn, P. 1996, A\&AS, 120, 667.
\reference{Ling et al. 1997}Ling, J. C., et al. 1997, ApJ, 484, 375
\reference{Ling et al. 2000}Ling, J. C., et al. 2000, ApJS, 127, 79
\reference{Masetti et al. 1996}Masetti, N., Bianchini, A., Bonibaker, J., Della Valle, M., $\&$ Viio, R., 1996, A\&A, 314, 123.4\
\reference{Mirabel et al. 1993}Mirabel, F., N., Rodriguez, L. F., $\&$ Cordier, B., 1993 IAU Circ 5876.
\reference{McConnell, et al. 2000}McConnell, M. L., et al. 2000, ApJ, 543, 928.
\reference{McConnell, et al. 2002}McConnell, M. L., et al. 2002, ApJ,  572,: 984.
\reference{Meier 2001}Meier, D. L., 2001, ApJ, 548, L9.
\reference{Meier 2004}Meier, D. L., 2004, private communication.
\reference{Phlips et al.1996}Phlips, B., et al., 1996, ApJ, 465, 907.
\reference{Poutanen \& Coppi, 1998}Poutanen, J. \& Coppi, P. S. 1998, Physica Scripta, T77, 57 (astro-ph/9711316).
\reference{Revnivtsev et al. 1998} Revnivtsev, J. P., et al. 1998, A\&A, 331, 557
\reference{Shakura $\&$ Sunyaev 1976}Shakura, N.I. $\&$ Sunyaev, R. A. 1976, MNRAS, 175, 613.
\reference{Sunyaev & Titarchuk 1980}Sunyaev , R. A., \& Titarchuk, L. G., 1980, A\&A, 86, 121.
\reference{Turolla et al. 2002}Turolla, R., Zane, S., \& Titarchuk, L., 2002, ApJ, 576, 349.
\reference{van der Hooft et al. 1996} van der Hooft, F., et al. 1996, ApJ, 458, L75.
\end{references}
\end{document}